\documentclass[final,3p,times]{elsarticle} 

\usepackage{lineno}
\usepackage{graphicx}
\usepackage{mathtools}
\usepackage{natbib}
\usepackage{xstring}
\usepackage{placeins}
\usepackage{amssymb}
\usepackage{amsmath}

\usepackage{color}      
\usepackage[colorlinks=true,
            linkcolor=BrickRed,
            urlcolor=blue,
            citecolor=blue]{hyperref}
            
\usepackage{multicol}%
\usepackage{comment}
\usepackage{lipsum}
\usepackage{amsthm}

\usepackage[figurename=Fig.]{caption}
\usepackage[abs]{overpic}    

\usepackage{subfig} 
\usepackage{float} 

\biboptions{sort&compress}

\journal{} 

\begin{document}

\begin{frontmatter}

\title{Model-form uncertainty quantification in RANS simulations \\ of wakes and power losses in wind farms}

\author[mysecondaddress]{Ali Eidi}
\author[mysecondaddress]{Reza Ghiassi}
\author[mythirdaddress]{Xiang Yang}
\author[mymainaddress]{Mahdi Abkar\corref{mycorrespondingauthor}}

\cortext[mycorrespondingauthor]{Corresponding author}
\ead{abkar@mpe.au.dk}
\address[mysecondaddress]{School of Civil Engineering, College of Engineering, University of Tehran, Tehran, Iran}
\address[mythirdaddress]{Department of Mechanical Engineering, Pennsylvania State University, State College, PA, 16802, USA}
\address[mymainaddress]{Department of Mechanical and Production Engineering, Aarhus University, 8000 Aarhus C, Denmark}

\begin{abstract}
Reynolds-averaged Navier-Stokes (RANS) is one of the most cost-efficient approaches to simulate wind-farm-atmosphere interactions. 
However, the applicability of RANS-based methods is always limited by the accuracy of turbulence closure models, which introduce various uncertainties into the models. 
In this study, we estimate model-form uncertainties in RANS simulations of wind farms. 
For this purpose, we compare different RANS models to a large-eddy simulation (LES). 
We find that the realizable $k-\varepsilon$ model is a representative RANS model for predicting the mean velocity, the turbulence intensity, and the power losses within the wind farm. 
We then investigate the model-form uncertainty associated with this turbulence model by perturbing the Reynolds stress tensor.  
The focus is placed on perturbing the shape of the tensor represented by its eigenvalues. 
The results show that the perturbed RANS model successfully estimates the region bounding the LES results for quantities of interest (QoIs). 
We also discuss the effect of perturbation magnitude on various QoIs.
\end{abstract}


\end{frontmatter}

\section{Introduction} \label{sec:Introduction}
Wind power has been one of the fastest-growing sustainable energy sources in the past few decades and has contributed to the ever-increasing efforts to mitigate greenhouse gas emissions and global warming. 
Increasing this contribution motivates more attention to larger and more efficient wind farms. 
This directly leads to computational fluid dynamics (CFD).
CFD is a cost-effective approach for modeling and analyzing wind farms, and it can provide detailed information about the wind-turbines-atmosphere interactions. 
However, it is known that inaccurate prediction of turbulent flow patterns -- particularly the wakes formed behind wind turbines -- significantly impact CFD's estimates of wind farms' power generation (see, e.g.,  the review of Refs. \cite{goccmen2016wind,Stevens2017,Archer2018}). 
This justifies efforts to increase the accuracy of the CFD models, on the one hand, and quantifying their uncertainties, on the other hand. 
In fact, standard CFD approaches are inherently uncertain, and therefore interpretation of any CFD simulation results must be associated with some level of uncertainty quantification (UQ) \cite{duraisamy2019turbulence,xiao2019quantification}. 

Different computational approaches have been considered by researchers and engineers, among which eddy-resolving simulations -- direct numerical simulation (DNS) and large-eddy simulation (LES) -- provide a high level of physical fidelity \cite{Pope2000}. However, these methods are computationally expensive at high Reynolds numbers, and hence, are more suitable for small-scale problems in fundamental research (see the review of Refs. \cite{Mehta2014,Sorensen2015,porte2020wind, yang2021grid}, and references cited therein). 
Consequently, Reynolds-averaged Navier-Stokes (RANS), a non-scale-resolving tool, is considered the standard method in industrial applications \cite{vermeer2003wind,sanderse2011review}. 
The reduced computational cost of RANS models is associated with the use of RANS closures, which fall short in a variety of physical scenarios, leading to uncertainties and errors in the predicted quantities of interest (QoIs) \cite{duraisamy2019turbulence,xiao2019quantification}.
Specifically, in the wind-energy community, it has been reported in several research publications that standard RANS models fail to accurately predict the wake flow and power loses in wind farms  \cite{rethore2009wind,cabezon11,prospathopoulos2011evaluation,antonini2019improving}, whereas eddy-resolving numerical tools such as LES compare well with laboratory and field-scale experiments \cite{troldborg2011numerical,Porte-Agel2011,Wu2015,stevens2018comparison,bossuyt2018large,laan2019power,yang2021high,ge2021large,zhang2021new,ge2020study}. 

In RANS models, the proportionality between the eddy-viscosity and the mean rate of strain, i.e., Boussinesq’s hypothesis, is one of the main sources of uncertainty \cite{duraisamy2019turbulence,xiao2019quantification}. 
Since such uncertainty arises due to the modeling assumptions, it is called the model-from (or structural) uncertainly. 
No matter how much the model constants of these models are tuned according to the problem, this structural uncertainty is indispensable, making it essential to quantify and understand this type of uncertainty in a RANS calculation. 
One way to quantify the model-form uncertainty in RANS is to decompose the Reynolds stress anisotropy tensor to its amplitude, shape, and orientation represented by turbulent kinetic energy (TKE), eigenvalues, and eigenvectors, respectively. 
By perturbing any of these values, one can assess the model-form-induced uncertainty in the solution. 
This method, which was first introduced by \citet{emory2013modeling}, has been applied to different flow problems, including wind engineering flows, urban canopy flows, streamlined surface flows \cite{gorle2015quantifying,garcia2017quantifying,cremades2019reynolds,gorle2019epistemic}. 
In wind-energy applications, this approach was recently utilized in the RANS simulation of a stand-alone wind-turbine wake \cite{hornshoj2021quantifying}. 
This paper extends the previous work, and we study the model-form uncertainty in wind-farm simulations.
Like the previous work that applies the UQ methodology in Ref. \cite{emory2013modeling} to different flow problems, the novelty of this paper does not lie in the methodology itself but rather in the uncertainty estimate.
As will be clear in the later sections, many of the results we obtain for a wind farm cannot be directly inferred from the results of a single turbine (as different physics is at play in these two flow configurations). 

The specific objective of the present work is to quantify the model-form uncertainty in RANS simulations of wind farms and the associated power losses. 
The QoIs are the mean velocity deficit, the turbulence intensity, and the turbine power output in wind farms. 
We will show that the propagation of uncertainties in a wind farm is very similar to that of turbine wakes: they propagate downstream, superimpose, and asymptote shortly after a few rows of turbines. 
The rest of the paper is organized as follows: \autoref{sec:math} introduces the RANS framework and a description of the numerical setup. A brief description of the LES code is also provided. In \autoref{sec:UQ}, we describe the UQ framework, and in \autoref{sec:Results}, we show the results. 
We compare different RANS models to LES and discuss the UQ results.
Finally, a summary and concluding remarks are given \autoref{sec:Conclusion}. 

\section{Mathematical details and simulation framework} \label{sec:math}

\subsection{Reynolds-averaged Navier-Stokes (RANS) framework} \label{sec:RANSf}
The RANS framework used in the present study solves the Reynolds-averaged version of continuity and momentum conservation equations for an incompressible flow,  

\begin{equation}
    \begin{gathered}
        \label{eq:intro-1}
        \partial_i \bar{u}_i = 0,
        \\        
         \partial_t \bar{u}_i
        + \bar{u}_j \partial_j \bar{u}_i 
        = -\frac{1}{\rho} \partial_i \bar{p}
        + \partial_j \left(2\nu \Bar{S}_{ij}-\overline{u_i'u_j'}\right)
        + f_i,
    \end{gathered}
\end{equation}
where the overbar and prime denote the Reynolds-averaged component and fluctuating component of a variable, respectively, $u_i$ is the velocity, $\rho$ is the density, $p$ is the pressure, $t$ is the time, $\nu$ is the molecular viscosity, $\bar{S}_{ij}$ is the mean rate of strain, $\overline{u_i'u_j'}$ is the Reynolds stress, and $f_i$ is a body force term representing the wind-turbine effects on the flow field. According to Boussinesq's hypothesis, the deviatoric part of the Reynolds stress tensor is linearly related to the mean rate-of-strain tensor as   

\begin{equation}
    \begin{gathered}
        \label{eq:intro-2}
         \overline{u_i'u_j'}-\frac{2}{3}k \delta_{ij}
        = - 2\nu_T \Bar{S}_{ij},
    \end{gathered}
\end{equation}
where ${k=\overline{u_i'u_i'}}/2$, ${\delta_{ij}}$ and ${\nu_{T}}$  represent TKE, Kronecker delta, and eddy viscosity, respectively. 

\begin{figure}
    \centering
    \includegraphics[width=0.6\linewidth]{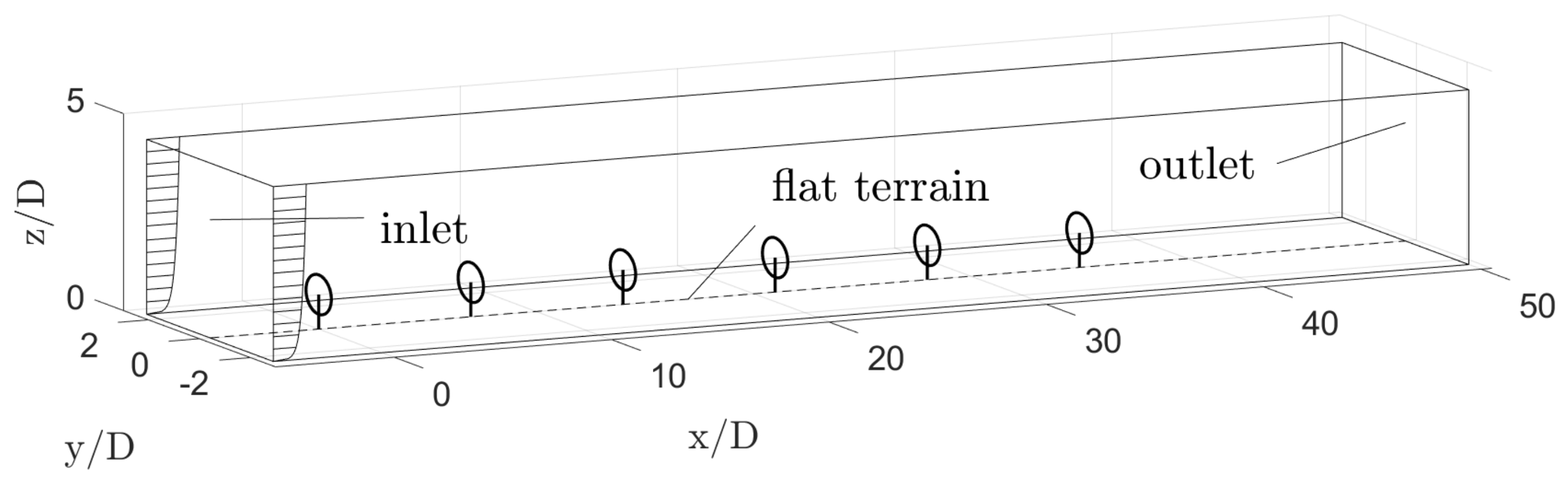}
    \caption{A schematic of the RANS computational domain.}
    \label{fig:domain}
\end{figure}

The computational domain used in RANS simulations is shown in \autoref{fig:domain}, and its size is set to ${4400\text{m} \times400\text{m} \times355\text{m}}$ in \emph{x}, \emph{y}, and \emph{z} directions, respectively. 
In the baseline case, the first turbine is located at a distance of five rotor diameters (\emph{D}) from the inlet, and the other turbines are located at a 7\emph{D} distance from the upstream turbine in an aligned configuration. 
This turbine spacing is quite typical in wind-farm implementations \cite{Meyers2012}, and the aligned configuration promotes wake interactions, whose effect on model uncertainty is a topic of this paper.
In addition to the baseline configuration, two additional turbine layouts with staggered turbine arrangement and closer turbine positioning, i.e., 5D will be investigated.  
The turbines have a diameter (\emph{D}) of 80m, a hub height (\emph{z$_{h}$}) of 70m, and an axial induction factor of ($a$) of 0.25. 
The turbine-induced forces are modeled using the standard non-rotational actuator-disk model described by \citet{Calaf2010} using the wind velocity at the rotor plane and the disk-based thrust coefficient, $C_T'=4a/(1-a)$, which is set to $4/3$.   

The inlet and top boundary conditions are defined by prescribing velocity, turbulence kinetic energy, and turbulence dissipation rate (or specific dissipation rate), corresponding to a logarithmic boundary-layer flow, together with a zero-gradient condition for the pressure \cite{van2019improved}.
The outflow boundary condition is a pressure outlet with zero gradients for the velocity and other turbulence quantities. 
A cyclic boundary condition is applied in the lateral direction for all the variables. A rough wall boundary condition with wall functions for the turbulence quantities is imposed at the bottom wall \cite{antonini2019improving}.
The inlet's velocity at the hub height ($\bar{u}_{h}$) is 8m/s, and the turbulence intensity (${I=\sqrt{2k/3}/\bar{u}_{h}}$) at the same height is 5.8\%. The aerodynamics surface roughness (\emph{z$_0$}) is determined for each turbulence model to satisfy the desired conditions at the hub height. The flow's governing equations are discretized and solved using the OpenFOAM open-source software, based on the finite volume approach and the SIMPLE (Semi-Implicit Method for Pressure-Linked Equations) algorithm \cite{ferziger2002computational}. Different turbulence models considered in this study are described in the next section. A grid convergence study is conducted, and the results are shown in \ref{app:appendixA}. 

\subsection{Large-eddy simulation (LES) framework} \label{sec:LES}
Here, an in-house pseudo-spectral code is employed for the LESs, which has been well validated and used in earlier wind-energy research publications (e.g., Refs. \cite{Wu2013,Abkar2014,Abkar2016,abkar2019theoretical}). 
The size of the computational domain is ${4800\text{m} \times 800\text{m}\times 355\text{m}}$ in \emph{x}, \emph{y}, and \emph{z} directions, respectively, and it is discretized uniformly into ${480 \times 160 \times 72}$ grid points. The LES domain in the \emph{x}-direction is slightly larger than the RANS setup due to the presence of a fringe zone, which is used to adjust the flow from the downstream wake state to an undisturbed inflow condition \cite{stevens2014concurrent}. 
There are two columns of six turbines with the same streamwise spacing as in the RANS simulations and with the lateral spacing of $5D$. 
The width of the RANS domain is half of the LES domain, and LES results are averaged over the two columns to be compared to RANS. 
For the sake of brevity, further details about the LES code have been excluded, as well as its equations here. A more detailed description of the LES code can be found in Refs. \cite{Abkar2015,Yang2018}. 

\section{Uncertainty quantification (UQ)} \label{sec:UQ}

The primary purpose of this study is to investigate the model-form uncertainty in our RANS simulations.
In the following, we describe the UQ methodology.
We represent the Reynolds stress tensor in terms of its amplitude, shape, and orientation \cite{emory2013modeling,thompson2019eigenvector}. The Reynolds stress anisotropy tensor (${b_{ij}}$), which is the deviatoric part of the Reynolds stress normalized by the TKE, can be written as \cite{banerjee2007presentation}  

\begin{equation}
    \begin{gathered}
        \label{eq:intro-3}
        b_{ij}
        =  \frac{\overline{u_i'u_j'}}{2k} 
        - \frac{1}{3} \delta_{ij}.
    \end{gathered}
\end{equation}
\noindent
The eigendecomposition of the normalized anisotropy tensor yields (see, e.g., Ref. \cite{kreyszig2009advanced})

\begin{equation}
    \begin{gathered}
        \label{eq:intro-4}
        b_{ij} 
        =  v_{im}{\Lambda}_{mn}v_{nj}.
    \end{gathered}
\end{equation}
Here, ${\Lambda}_{mn}$ is a diagonal tensor containing the eigenvalues of the anisotropy tensor in a decsending order (i.e., ${\lambda_{1}} \geq \lambda_{2} \geq {\lambda_{3}}$), and $v_{ij}$ is a tensor containing the orthogonal eigenvectors of the anisotropy tensor.
Note that due to the normalization constraints, the trace of the diagonal eigenvalues tensor is equal to zero, i.e., ${\lambda_{1} + \lambda_{2} + \lambda_{3} = 0}$ \cite{banerjee2007presentation}.
Using \autoref{eq:intro-3} and \autoref{eq:intro-4}, the Reynolds stress tensor can be written as

\begin{equation}
    \begin{gathered}
        \label{eq:intro-5}
         \overline{u_i'u_j'}
        = {2k} \left(\frac{1}{3}\delta_{ij}
        + v_{im}\Lambda_{mn}v_{nj}\right).
    \end{gathered}
\end{equation}
\noindent
Applying perturbations to the amplitude ($k$), shape ($\mathbf{\Lambda}$), and orientation ($\mathbf{v}$) of this tensor, we have the following perturbed Reynolds stress tensor  

\begin{equation}
    \begin{gathered}
        \label{eq:intro-6}
         \overline{u_i'u_j'}^*
        = {2k^*} \left(\frac{1}{3}\delta_{ij}
        + v^*_{im}\Lambda^*_{mn}v^*_{nj}\right),
    \end{gathered}
\end{equation}
where ${k^*}$, ${v_{im}^*}$ and ${\Lambda^*_{mn}}$ are the perturbed TKE, eigenvectors and eigenvalues, respectively.
Adding the difference between the perturbed and the original Reynolds stress tensors 
(i.e., $\Delta R_{ij}=\overline{u_i'u_j'}^*-\overline{u_i'u_j'}$) to the RANS equations gives rise to the perturbed RANS equations 

\begin{equation}
    \begin{gathered}
        \label{eq:intro-8}
         \partial_i \bar{u}_i = 0,     \\
         \partial_t \bar{u}_i
        + \bar{u}_j \partial_j \bar{u}_i 
        = -\frac{1}{\rho} \partial_i \left(\bar{p}+\frac{2}{3}\rho k \right)
        + \partial_j \left[2(\nu+\nu_T) \Bar{S}_{ij}\right]
        + f_i 
        -\partial_j \Delta R_{ij}.        
    \end{gathered}
\end{equation}
$\Delta R_{ij}$ is the model-form uncertainty.
Following Refs. \cite{cremades2019reynolds,hornshoj2021quantifying}, we implement \autoref{eq:intro-8} in the open-source solver OpenFOAM. 
Here, we focus on perturbing the eigenvalues of the anisotropy tensor, which can be represented on a Barycentric map. 
Having a triangle, each of its corners represents a limiting state of one-component ${\hat{a}_{1}}$, two-component ${\hat{a}_{2}}$, and three-component (or isotropic) ${\hat{a}_{3}}$ turbulence. One can show that any Reynolds stress can be represented on the Barycentric triangle map by a linear combination as  

\begin{equation}
    \begin{gathered}
        \label{eq:intro-9}
        \textbf{x}
        =  \hat{\emph{\textbf{a}}}_{1}(\lambda_1 - \lambda_2)
    +  \hat{\emph{\textbf{a}}}_{2}(2\lambda_2 - 2\lambda_3)
        +  \hat{\emph{\textbf{a}}}_{3}(3\lambda_3 +1 ),
    \end{gathered}
\end{equation}
\noindent where 

\begin{equation}
    \begin{gathered}
        \label{eq:intro-10}
        \hat{\emph{\textbf{a}}}_{1} = 
        \begin{bmatrix}
        1\\
        0
        \end{bmatrix},\ \hat{\emph{\textbf{a}}}_{2} =
        \begin{bmatrix}
        -1\\
        0
        \end{bmatrix}\ ,\ \hat{\emph{\textbf{a}}}_{3} =
        \begin{bmatrix}
        0\\
        \sqrt{3}
        \end{bmatrix},
    \end{gathered}
\end{equation}
\noindent are the coordinates of the Barycentric triangle's corners \cite{emory2013modeling}. The shape of the Reynolds stress tensor can be perturbed towards one-, two-, and three-component limiting states as 

\begin{equation}
    \begin{gathered}
        \label{eq:intro-11}
            \bf{\Lambda}^*= (1-\delta) \bf{\Lambda}+ \delta \bf{\Lambda}_{c},\ 0 \leq \delta \leq 1, 
    \end{gathered}
\end{equation}
\noindent where ${\delta}$ is the amount of perturbation from 0 to 1, and  ${\bf{\Lambda}_{c}}$ could be one of the limiting states defined as 

\begin{equation}
    \begin{gathered}
        \label{eq:intro-12}
        {\bf{\Lambda}}_{1} = 
        \begin{bmatrix}
        2/3 & 0 & 0\\
        0 & -1/3 & 0\\
        0 & 0 & -1/3 
        \end{bmatrix},\ {\bf{\Lambda}}_{2} =
        \begin{bmatrix}
        1/6 & 0 & 0\\
        0 & 1/6 & 0\\
        0 & 0 & -1/3
        \end{bmatrix},\ {\bf{\Lambda}}_{3} =
        \begin{bmatrix}
        0 & 0 & 0\\
        0 & 0 & 0\\
        0 & 0 & 0
        \end{bmatrix}.
    \end{gathered}
\end{equation}

\begin{figure}
    \centering
    \includegraphics[width=0.6\linewidth]{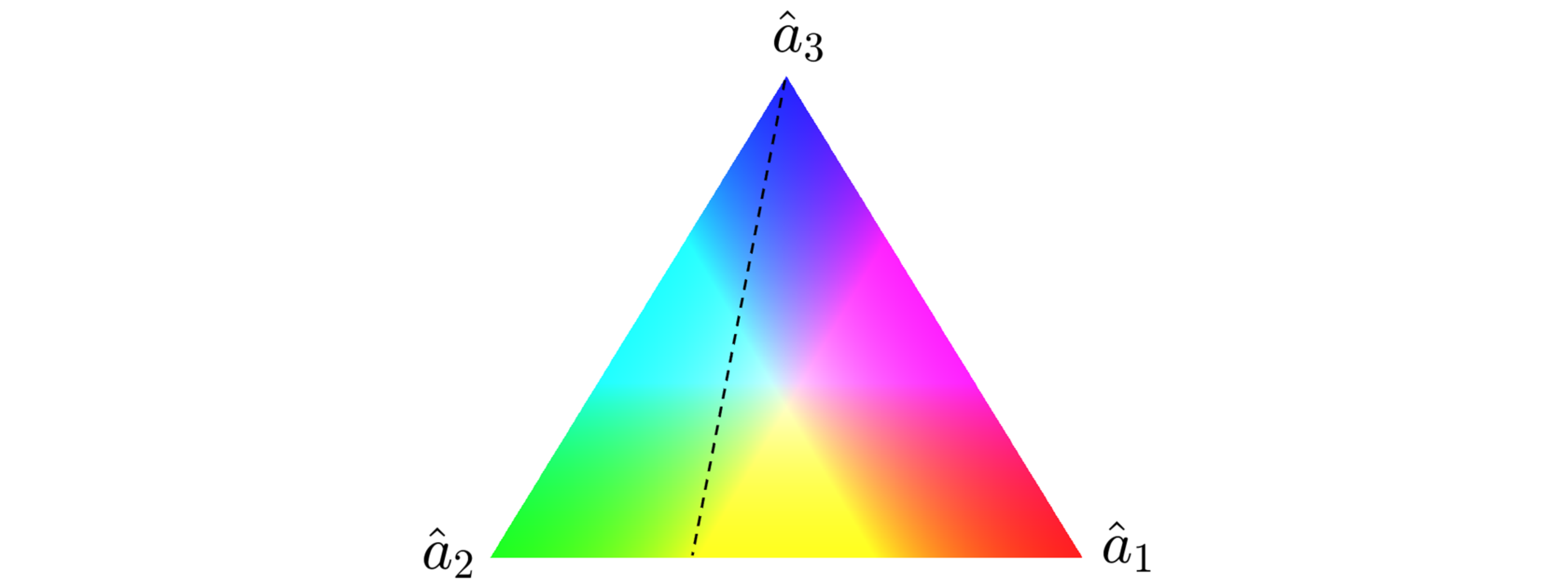}      
    \caption{Barycentric triangle colormap representation according to the convention in Ref. \cite{emory2014visualizing}. The dashed line corresponds to a plane shear flow, in which at least one of the eigenvalues of the Reynolds stress anisotropy tensor is zero.}
    \label{fig:RGBTri}
\end{figure}

It worth mentioning that by assigning an RGB color code to each of the points inside the Barycentric triangle, the Reynolds stress tensor can be visualized at any arbitrary location. The color interpretation used in this study is shown in \autoref{fig:RGBTri}. 
The following equation calculates the RGB code for any point in the Barycentric triangle map as 

\begin{equation}
    \begin{gathered}
        \label{eq:intro-13}
            \begin{bmatrix}
            R\\
            G\\
            B
            \end{bmatrix}
            = C_{1}^5 \begin{bmatrix}
            1\\
            0\\
            0\\ \end{bmatrix}
            +\ C_{2}^5 \begin{bmatrix}
            0\\
            1\\
            0 \end{bmatrix}
             +\ C_{3}^5 \begin{bmatrix}
            0\\
            0\\
            1 \end{bmatrix},
    \end{gathered}
\end{equation}
where $C_{1} = (\lambda_{1}-\lambda_{2})+0.65$, $C_{2} = 2(\lambda_{2}-\lambda_{3})+0.65$, and $C_{3} = (3\lambda_{3}+1)+0.65$ \cite{ali2019}. Here, the coefficients of 0.65 and 5 are for optimal visualization and better combination of colors. 

\section{Results and discussion} \label{sec:Results}

\subsection{Evaluation of different RANS models} \label{sec:RANSm}
Among different closure models that can be used to determine the Reynolds stress term, 
the two-equation eddy-viscosity models are the most commonly used by the wind-energy community, and this study focuses on two-equation models. 
In this study, we assess the performance of five different closure models: standard ${k-\varepsilon}$ \cite{jones1972prediction,launder1974application}, fp ${k-\varepsilon}$ \cite{van2015improved}, RNG ${k-\varepsilon}$ \cite{yakhot1992development}, realizable ${k-\varepsilon}$ \cite{shih1994new}, and SST ${k-\omega}$ \cite{menter1993zonal}. 
Note that a detailed comparison of RANS models is not the purpose of the present work as it has been addressed in the previous studies (e.g., Refs. \cite{cabezon2011comparison,simisiroglou14,hennen2017contribution,naderi2018modeling,antonini18}, among others). 
The purpose here is to select a baseline RANS model for the UQ study. 
The details of the turbulence models (their equations and model constants) are omitted for the sake of brevity, and the reader may refer to the above-mentioned references for further information.
In the following, the results of different RANS simulations are presented and compared with the LES data. The focus is placed on the prediction of normalized velocity deficit, defined as $\Delta \bar{u}/\bar{u}_h = (\bar{u}_{in}-\bar{u})/\bar{u}_h$, turbulence intensity, and power outputs in the wind farm. Here subscripts $in$ denotes inflow.

\begin{figure}
    \centering
    \includegraphics[width= 0.6\linewidth]{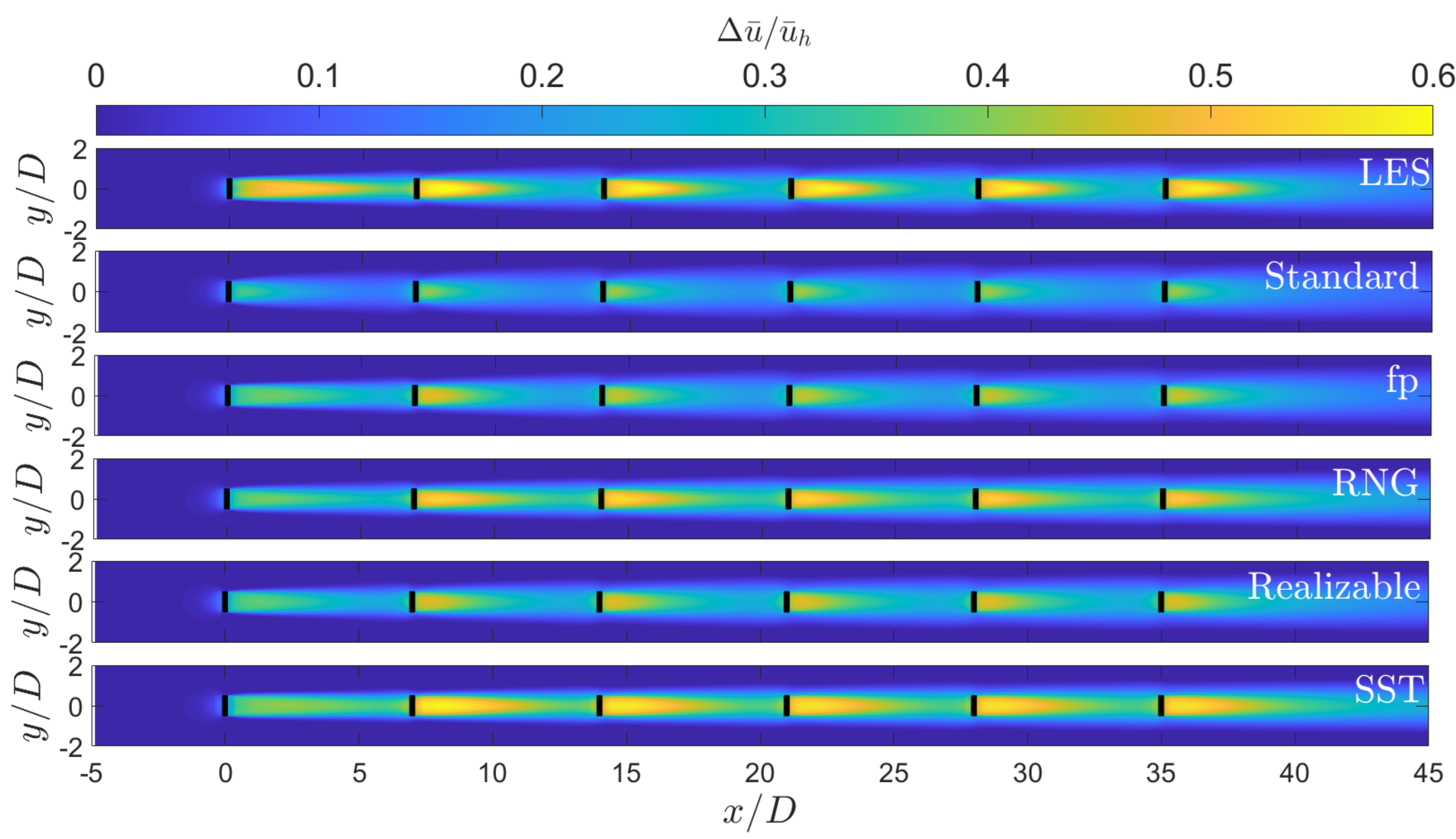}
    \caption{Normalized velocity deficit contour plots in a horizontal plane at the turbine hub height. From top to the bottom: LES, Standard k-$\varepsilon$, fp k-$\varepsilon$, RNG k-$\varepsilon$, Realizable k-$\varepsilon$ and SST k-$\omega$.}
    \label{fig:NVD_hub}
\end{figure}

\begin{figure}
    \centering
    \includegraphics[width=0.6 \linewidth]{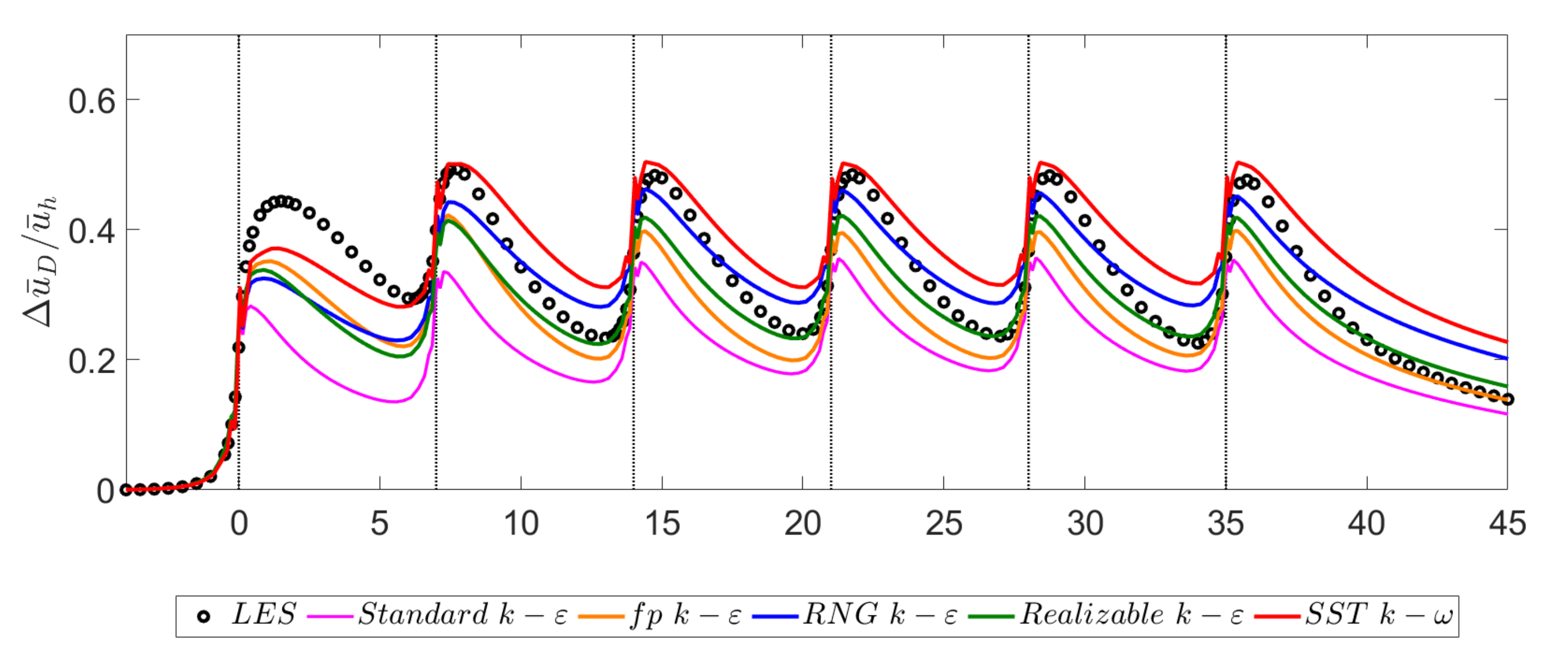}  \caption{Normalized velocity deficit, averaged over the rotor area, for different RANS closure models. Vertical dashed lines show the turbine locations.}
    \label{fig:averaged_NVD}
\end{figure}

The simulated normalized velocity deficit is shown in \autoref{fig:NVD_hub} in a two-dimensional horizontal plane at the turbine hub height. The normalized velocity deficit, averaged over the rotor area, is also plotted as a function of the downwind distance in \autoref{fig:averaged_NVD}. 
Both figures indicate that the RANS models considered here underestimate the wake velocity deficit behind the most upstream wind turbine. 
The SST k-$\omega$ model has a comparably better performance behind the first turbine, but its prediction deviates from the LES for the downstream waked wind turbines. The largest deviation, especially behind the first turbine, is observed for the standard k-$\varepsilon$, which can be attributed to the overestimation of the eddy viscosity immediately behind the turbine \cite{rethore2009wind,van2015improved}.  
We see from the two figures that, for the particular case considered here, the RNG k-$\varepsilon$, the fp k-$\varepsilon$, and the realizable k-$\varepsilon$ models show relatively good agreement with the LES data in the near wake and far wake for all downstream turbines.

\begin{figure}
    \centering
    \includegraphics[width= 0.6\linewidth]{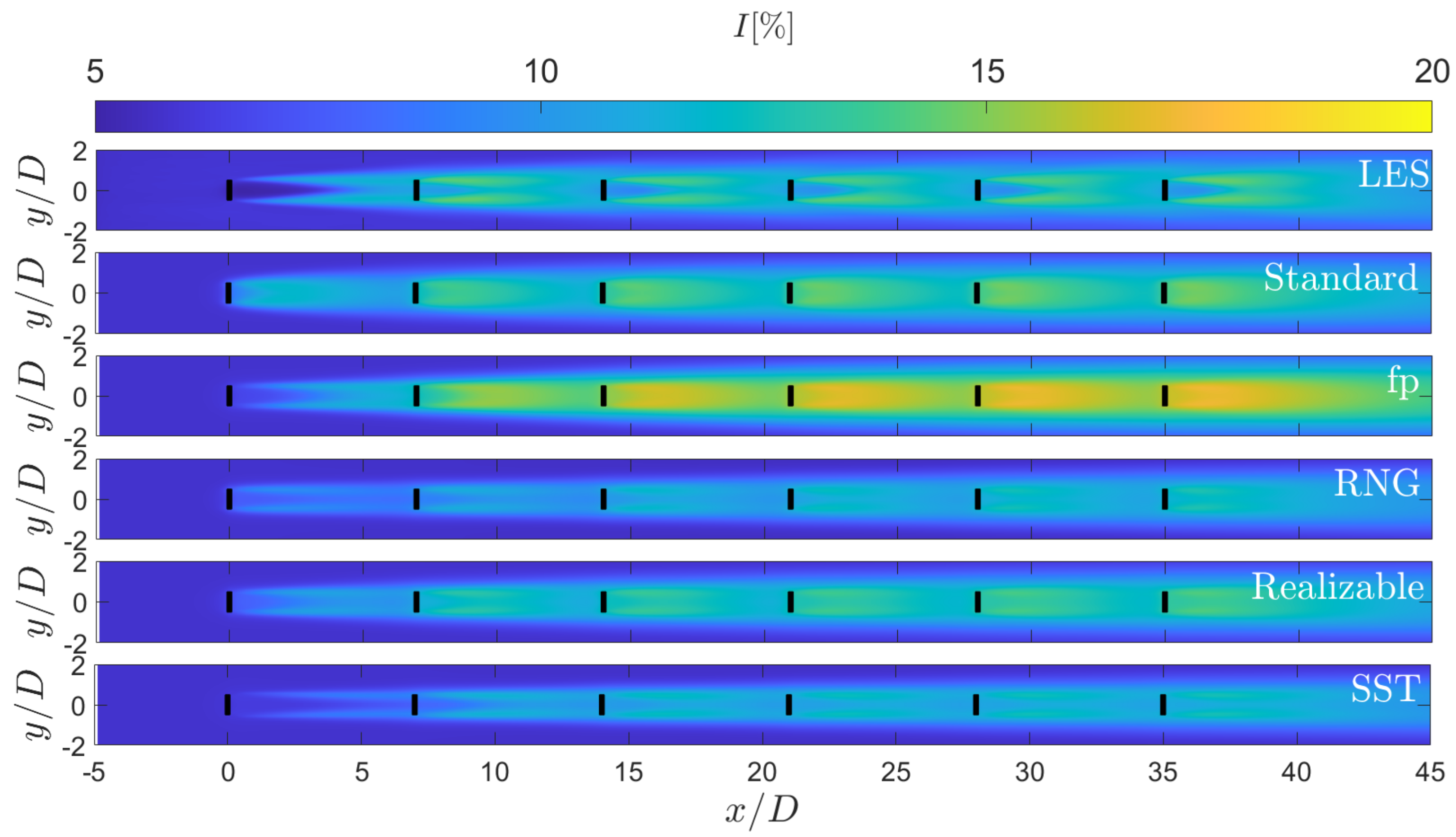}      
    \caption{Turbulence intensity contour plots in a horizontal plane at the turbine hub height. From top to the bottom: LES, Standard k-$\varepsilon$, fp k-$\varepsilon$, RNG k-$\varepsilon$, Realizable k-$\varepsilon$ and SST k-$\omega$.}
    \label{fig:I_hub}
\end{figure}

\begin{figure}
    \centering
    \includegraphics[width= 0.6\linewidth]{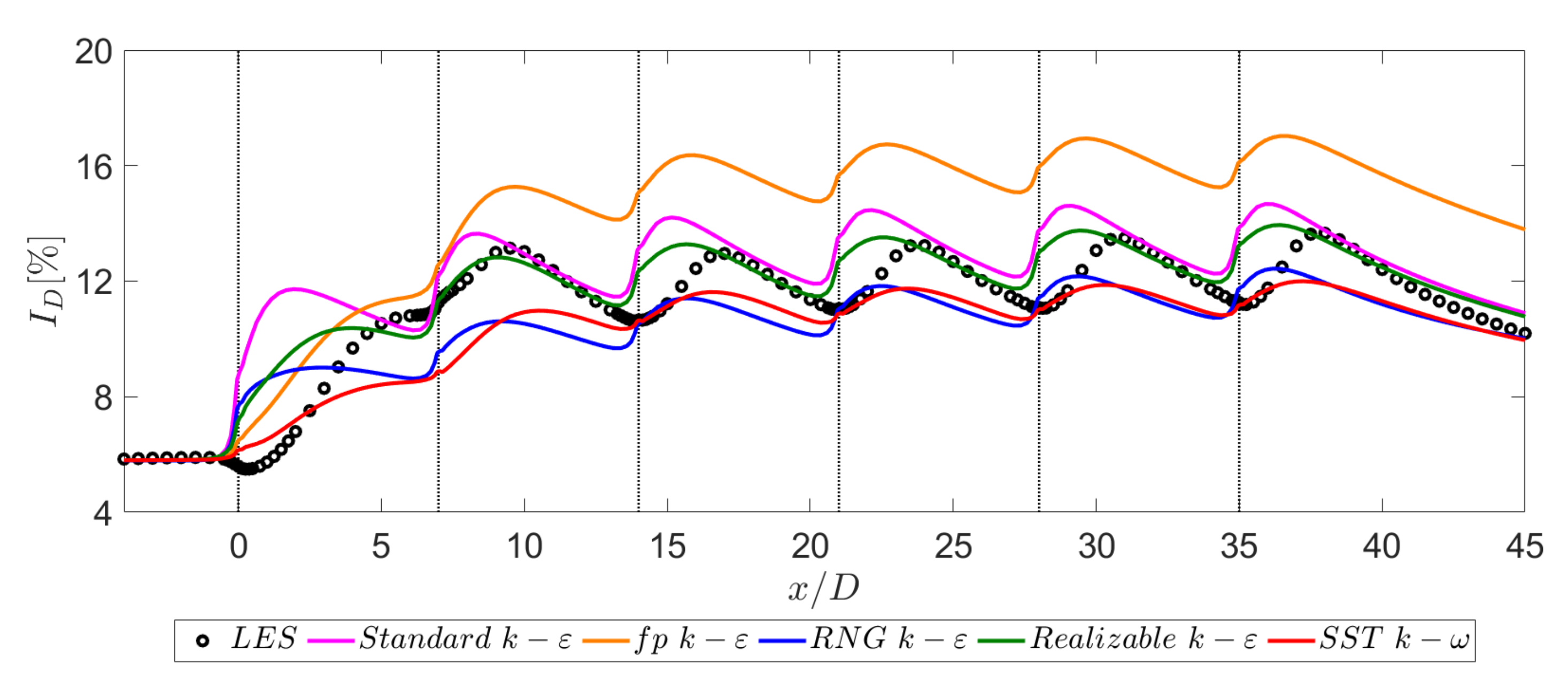}      
    \caption{Turbulence intensity, averaged over the rotor area, for different RANS closure models. Vertical dashed lines show the turbine rows.}
    \label{fig:averaged_I}
\end{figure}

\autoref{fig:I_hub} shows the contours of the turbulence intensity at turbine hub height. The turbulence intensity, averaged over the rotor area, is also plotted as a function of the downwind distance in \autoref{fig:averaged_I}. 
We see that the standard k-$\varepsilon$ does not do well in term of estimating the turbulence intensity in the wake of the first turbine. In particular, unlike the other RANS models, it fails to capture the double peak structure of the turbulence intensity behind the first turbine. The fp k-$\varepsilon$ performs well behind the first turbine, but it significantly overpredicts the turbulence intensity downstream of the other turbines. It can also be seen that the SST k-$\omega$ and RNG k-$\varepsilon$ models underestimate turbulence intensity for downstream waked turbines. 
Overall, we see that compared with other models, the Realizable k-$\varepsilon$ shows a better agreement with the LES data in estimating the turbulence intensity.

\begin{figure}
    \centering
    \includegraphics[width= 0.55\linewidth]{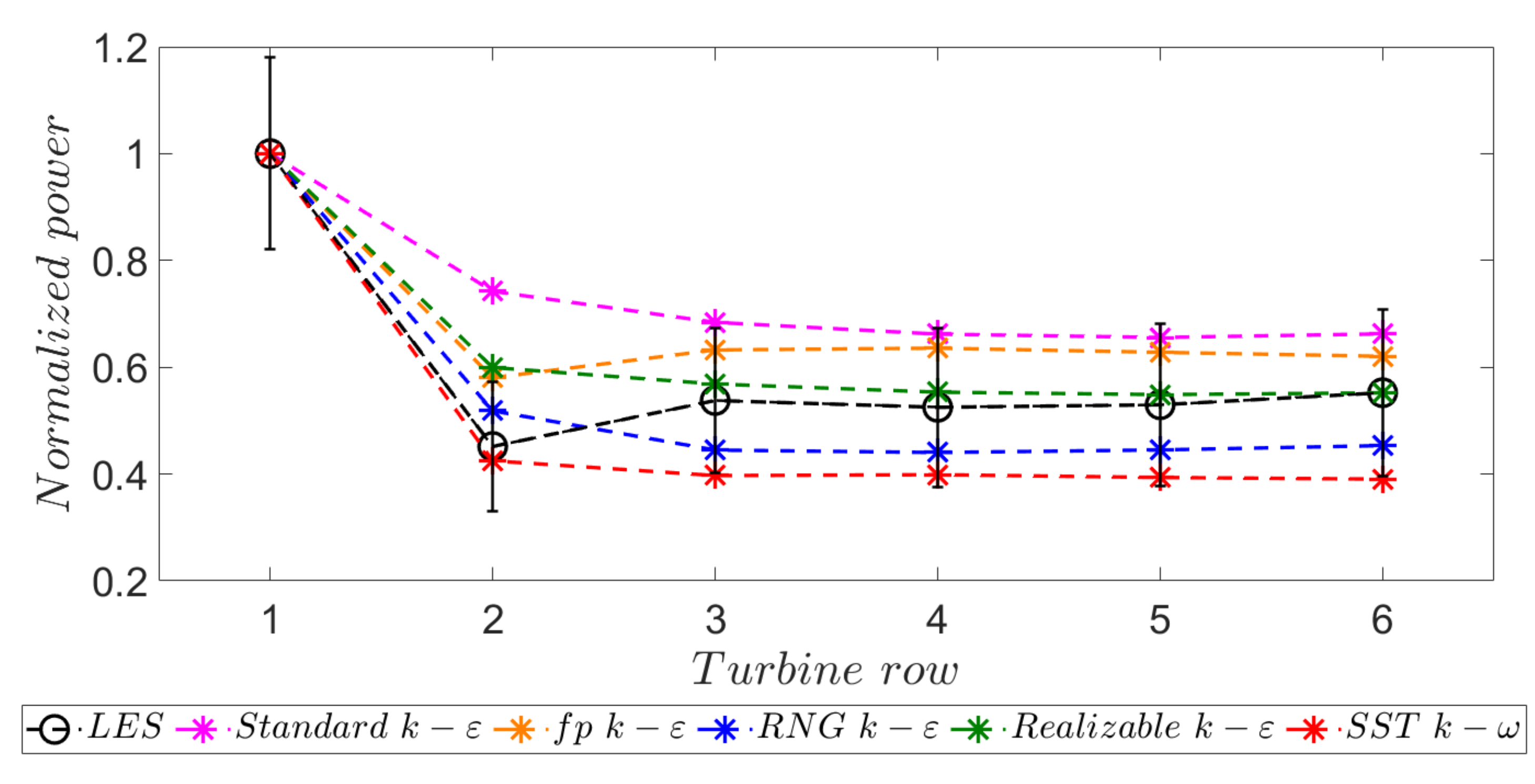}
    \caption{Normalized power output as a function of turbine row for different RANS closure models.}
    \label{fig:powers}
\end{figure}

A comparison of different RANS models in predicting wind-farm power output is provided in \autoref{fig:powers}. The results are normalized with the power of the most upstream wind turbine. Consistent with the results shown in \autoref{fig:NVD_hub} and \autoref{fig:averaged_NVD}, it can be seen that the standard k-$\varepsilon$ and the fp k-$\varepsilon$ overestimates the power output for the waked turbines. SST k-$\omega$ model provides a good estimation for the power output of the first downstream turbine but underestimates the power for the rest of the downstream turbines. RNG k-$\varepsilon$ model shows a similar trend as SST k-$\omega$ model.
The realizable k-$\varepsilon$ shows a better agreement with the LES in predicting the power output, although its prediction of the efficiency of the first downstream wind turbine is not entirely accurate. 

Overall, we find  the realizable k-$\varepsilon$ model performs well and is a representative RANS model. Hence, it is selected as the baseline model for the UQ study.

\subsection{Uncertainty quantification (UQ) analyses} \label{sec:UQa}

\begin{figure}
    \centering
    \includegraphics[width=0.6\linewidth]{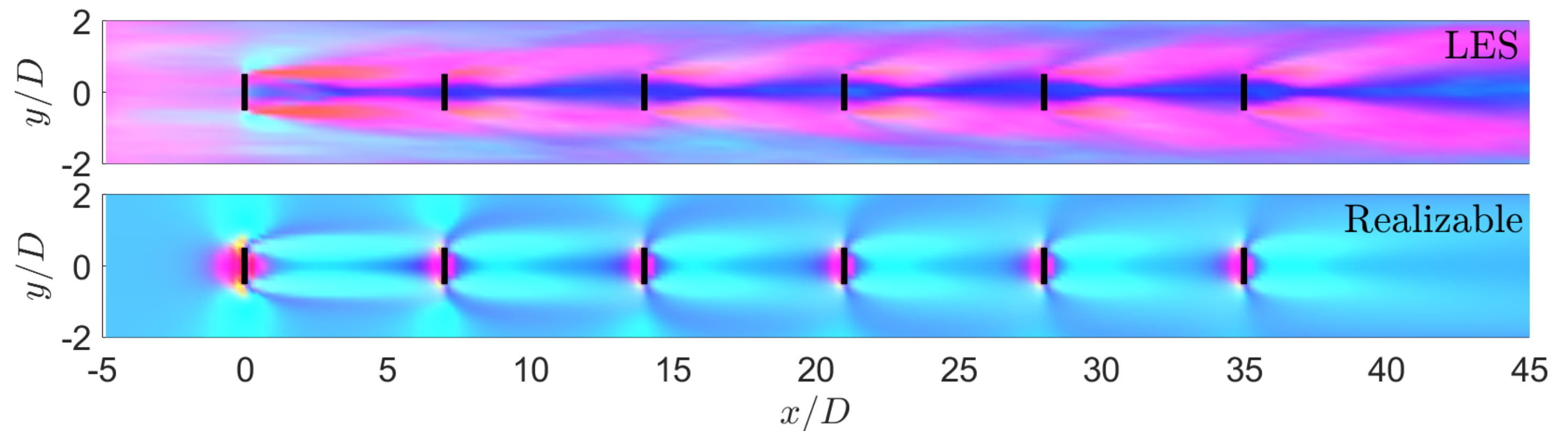}    
    \caption{Reynolds stress anisotropy tensor for LES (top) and RANS (bottom) simulations depicted in the Barycentric RGB map.}
    \label{fig:RGB}
\end{figure}

Before we proceed with detailed UQ analysis, we first examine the Reynolds stress anisotropy in both RANS and LES.
First, \autoref{fig:RGB} shows the RGB colormap of the Reynolds stress anisotropy tensor at the horizontal plane through the hub height for LES and the baseline RANS model following \autoref{eq:intro-13}. 
As can be seen, near the two side-tip positions, the Reynolds stress obtained from LES tends to be more anisotropic (towards the one-component turbulence) compared to the RANS results. This is expected as the RANS eddy-viscosity models underestimate the level of anisotropy in the regions with a high level of shear. It is also observed that, in the core of the wake, the baseline RANS model underestimates the level of isotropy compared to the LES data. 

\begin{figure}
    \centering
    \includegraphics[width= 0.6\linewidth]{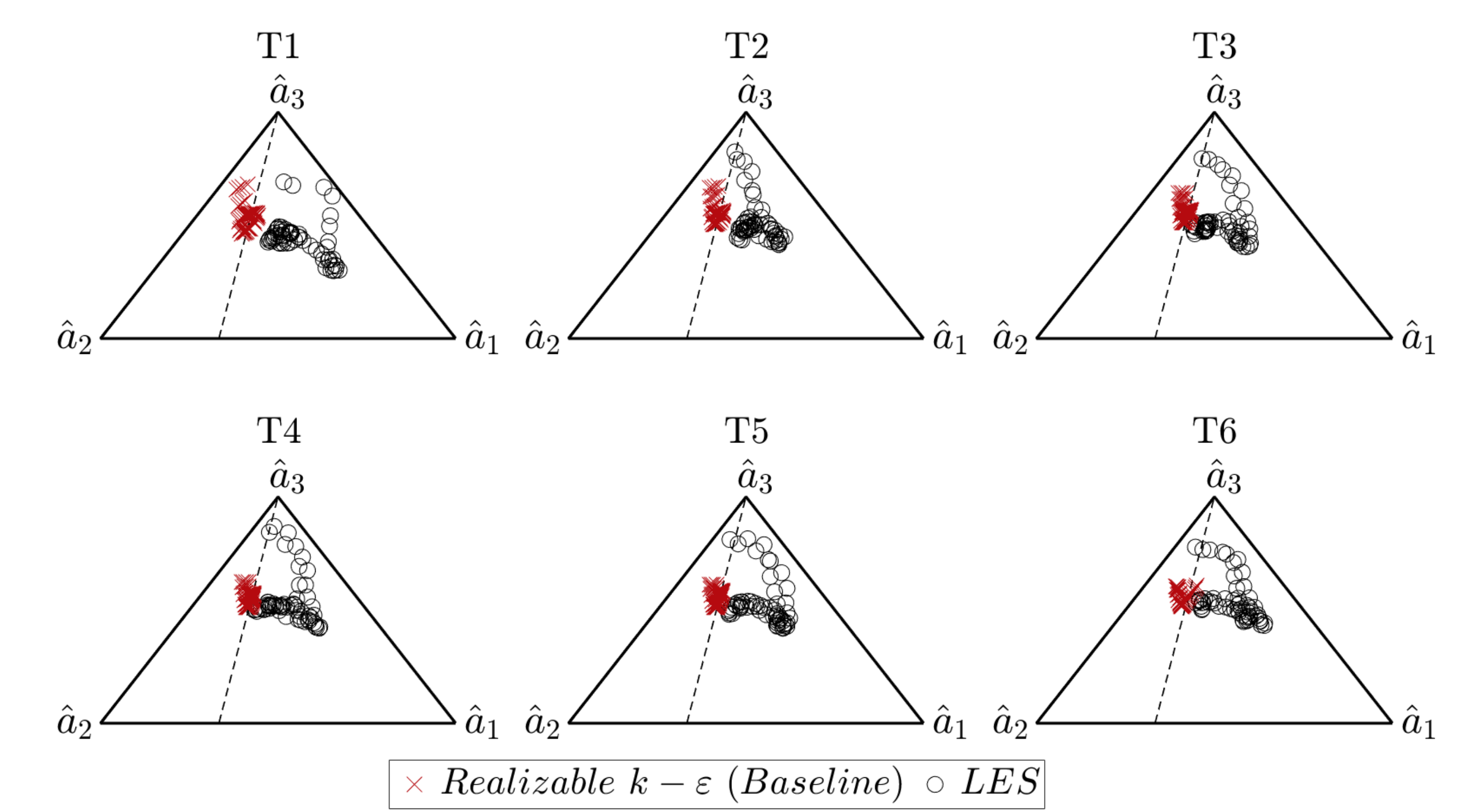}      
    \caption{Reynolds stress anisotropy tensor for the LES and RANS simulations at $x/D=5$ behind each turbine depicted in the Barycentric coordinate system.}
    \label{fig:Bary}
\end{figure}

To have a more quantitative insight on the results presented in \autoref{fig:RGB}, the structure or the Reynolds stress anisotropy on the Barycentric map is explored in \autoref{fig:Bary} at 5D downstream of each individual turbine and at the hub height level, for both LES and RANS models. The results show a fairly similar pattern for all six locations. In all cases, the RANS predictions are mostly around the plane strain line. In contrast, LES data are more inclined towards one-component and three-component turbulence. 

The results shown in \autoref{fig:Bary} indicate that a moderate perturbation of the Reynolds stress tensor towards one-component and three-component turbulence will make the RANS predicted Reynolds stress more similar to the LES under this particular metric, and that a perturbation towards the two-component extreme will make the RANS-predicted Reynolds stress less like the LES. 
However, a close agreement in the Barycentric map does not guarantee whatsoever a close agreement for other QoIs.
That is, one cannot infer from \autoref{fig:Bary} how a perturbation towards the one-, two-, three-component extremes will affect RANS's prediction of QoIs like the velocity deficit, the turbulence intensity, and the power loss. 
As will be shown later, perturbation of the RANS model toward the two-component turbulence leads to a prediction that lies within the ones obtained from the other two limiting states of turbulence. 
Here, three different values for the magnitude of perturbation $\delta$ are considered, i.e, 0.25, 0.5, and 1. The $\delta=0.5$ is adopted based on \textit{a priori} studies \cite{hornshoj2021quantifying}, which is also in good agreement with the results presented here. 
In addition to this value, one extremely conservative ($\delta=1$) and one less conservative value ($\delta=0.25$) are also considered to analyze the effect of injected perturbation amount on the QoIs.     
Here, we follow the previous work and use a uniform $\delta$.
We can get a non-uniform $\delta$ by make it a random number, which may not be very meaningful.
A more meaningful way of getting a non-uniform $\delta$ is to make it a function of the local flow condition \cite{gorle2014deviation}.
This, however, is usually only attempted when calibrating a model, in which case, the function is obtained through fitting.
As model calibration falls out of the scope of this work, we follow the previous study and use a uniform $\delta$.

\begin{figure}
    \centering
    \includegraphics[width= 0.6\linewidth]{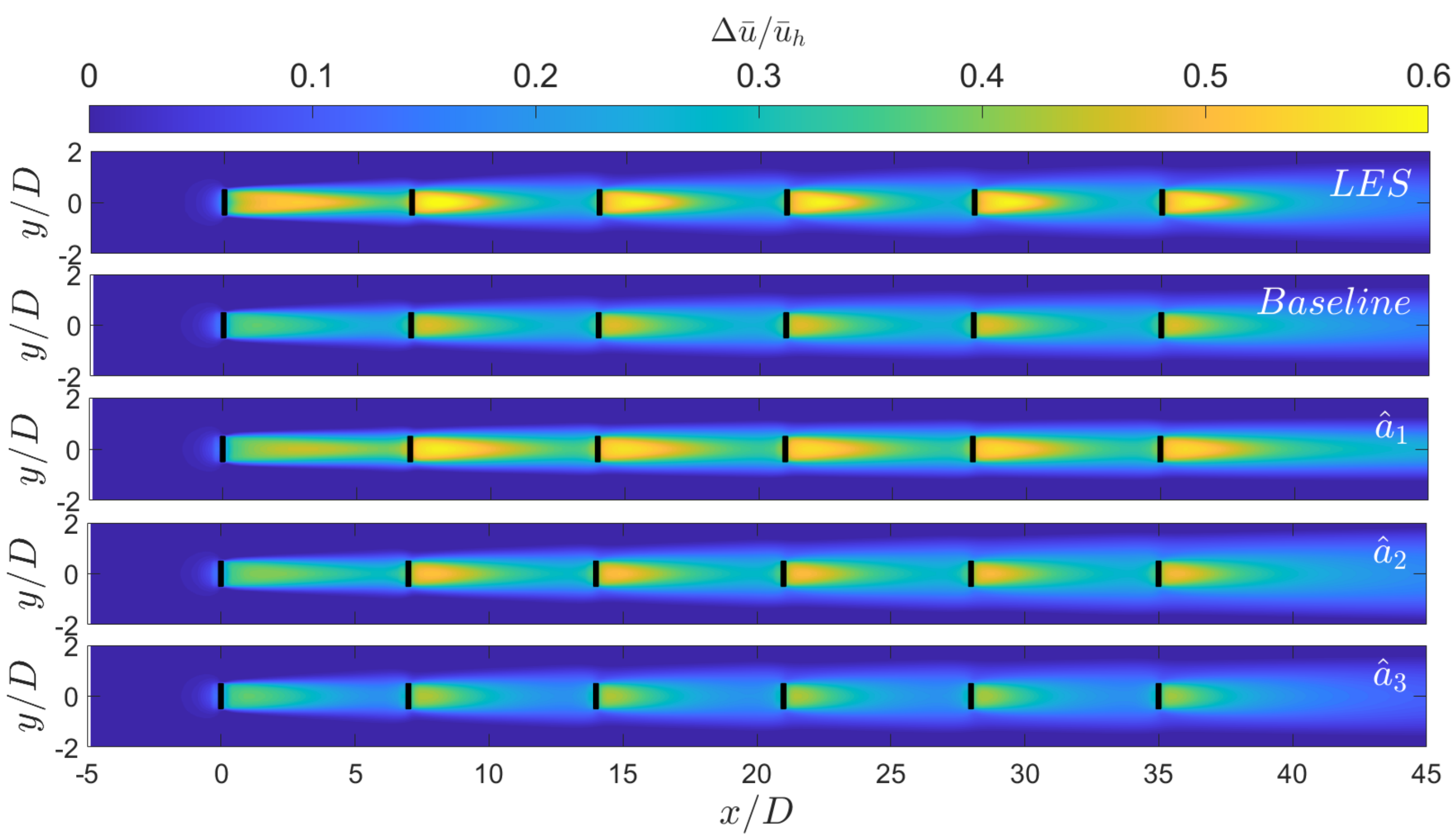}      
    \caption{Normalized velocity deficit contour plots in the horizontal plane at the turbine hub height. From top to the bottom: LES, Baseline RANS, and perturbed RANS towards one-, two- and three-component turbulence with $\delta$ = 0.5.}
    \label{fig:NVD0.5}
\end{figure}

\autoref{fig:NVD0.5} represents normalized velocity deficit contours for ${\delta = 0.5}$ at the horizontal plane through the hub height. As can be seen, perturbation towards one-component turbulence increases the velocity deficit, leading to a more extended wake. In contrast, perturbation toward the three-component (isotropic) state enhances the mixing and leads to a faster wake recovery. It is also observed that perturbation towards the two-component turbulence yields a prediction that lies within the ones obtained from the one-component and three-component states of turbulence. 

\begin{figure}
    \centering
    \includegraphics[width= 0.6\linewidth]{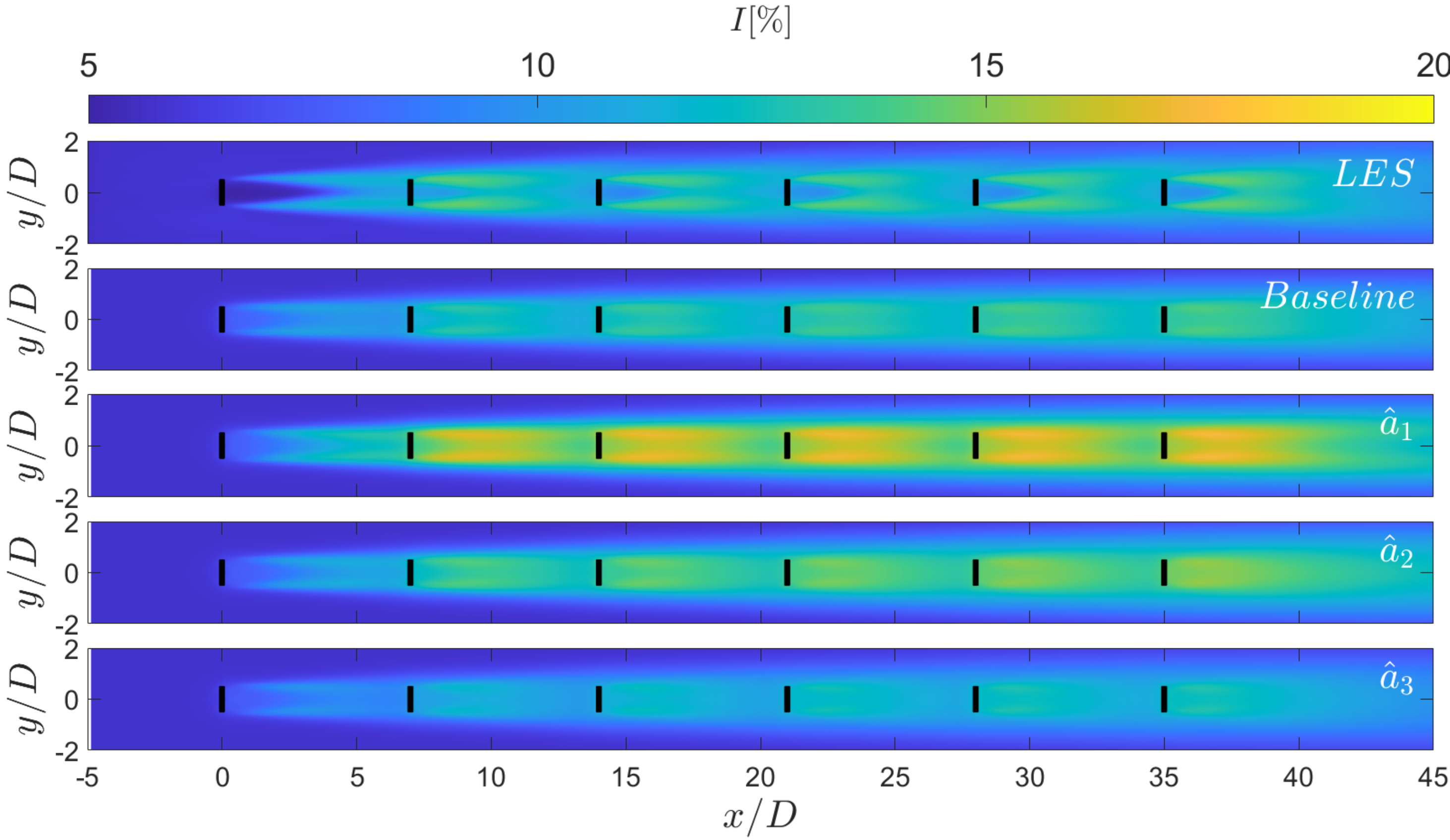}      
    \caption{Turbulence intensity contour plots in the horizontal plane at the turbine hub height. From top to the bottom: LES, Baseline RANS, and perturbed RANS towards one-, two- and three-component turbulence with $\delta$ = 0.5.}
    \label{fig:I0.5}
\end{figure}

\autoref{fig:I0.5} shows the effect of Reynolds stress perturbation towards three limiting states on the turbulence intensity for $\delta$ = 0.5. As expected, perturbation towards $\hat{a}_{1}$ increases the shear and, consequently, the turbulence intensity behind the turbines. In contrast, the perturbation toward the three-component state leads to a reduction in the turbulence level. These results are consistent with the ones obtained for the wake behind a stand-alone wind turbine \cite{hornshoj2021quantifying}. 
Similar to the results for normalized velocity deficit, model perturbation towards the two-component turbulence lies between the predictions obtained from one-component and three-component states. 

\begin{figure}
    \centering
    \includegraphics[width= 0.6\linewidth]{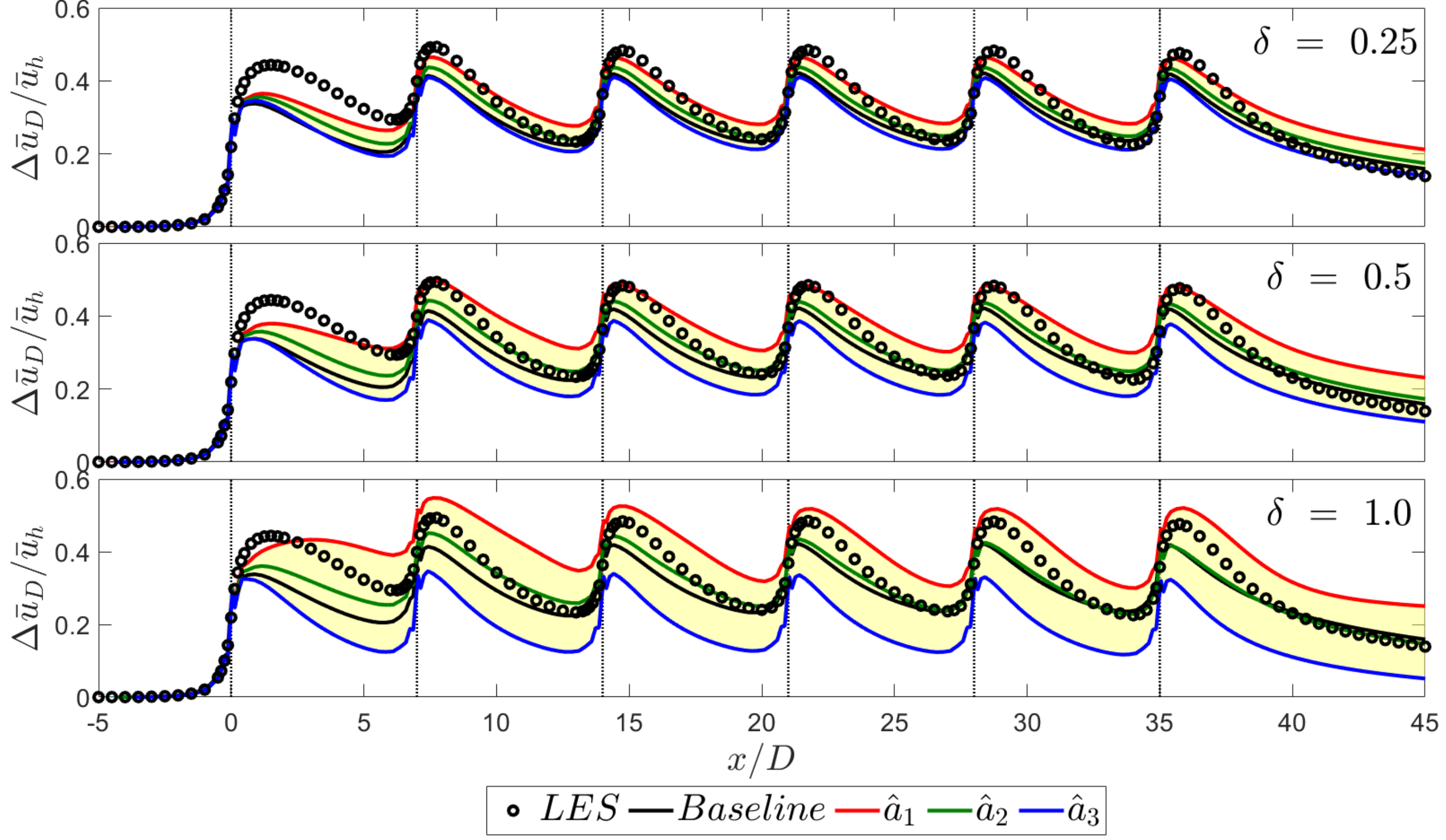}      
    \caption{Normalized velocity deficit, averaged over the rotor area, for LES, Baseline RANS, and perturbed RANS towards one-, two- and three-component turbulence with (from top to the bottom) $\delta = 0.25$, $\delta = 0.5$ and $\delta$ = 1.0.}
    \label{fig:perturbed_average_NVD}
\end{figure}

The evolution of the normalized velocity deficit, averaged over the rotor area, is shown in \autoref{fig:perturbed_average_NVD} for different amounts of perturbations. 
As can be observed, perturbation toward one-component turbulence increases the velocity deficit; perturbation toward the three-component state decreases the velocity deficit; and perturbation toward the two-component turbulence lies within the other two states. It is also evident that increasing the $\delta$ value from 0.25 to 0.5 and subsequently to 1.0 increases the uncertainty bands in the RANS prediction. It can be seen that, except in the near wake region behind the first turbine, the LES results are well covered in the band between the one-component and three-component curves in \autoref{fig:perturbed_average_NVD}.    

\begin{figure}
    \centering
    \includegraphics[width=0.6\linewidth]{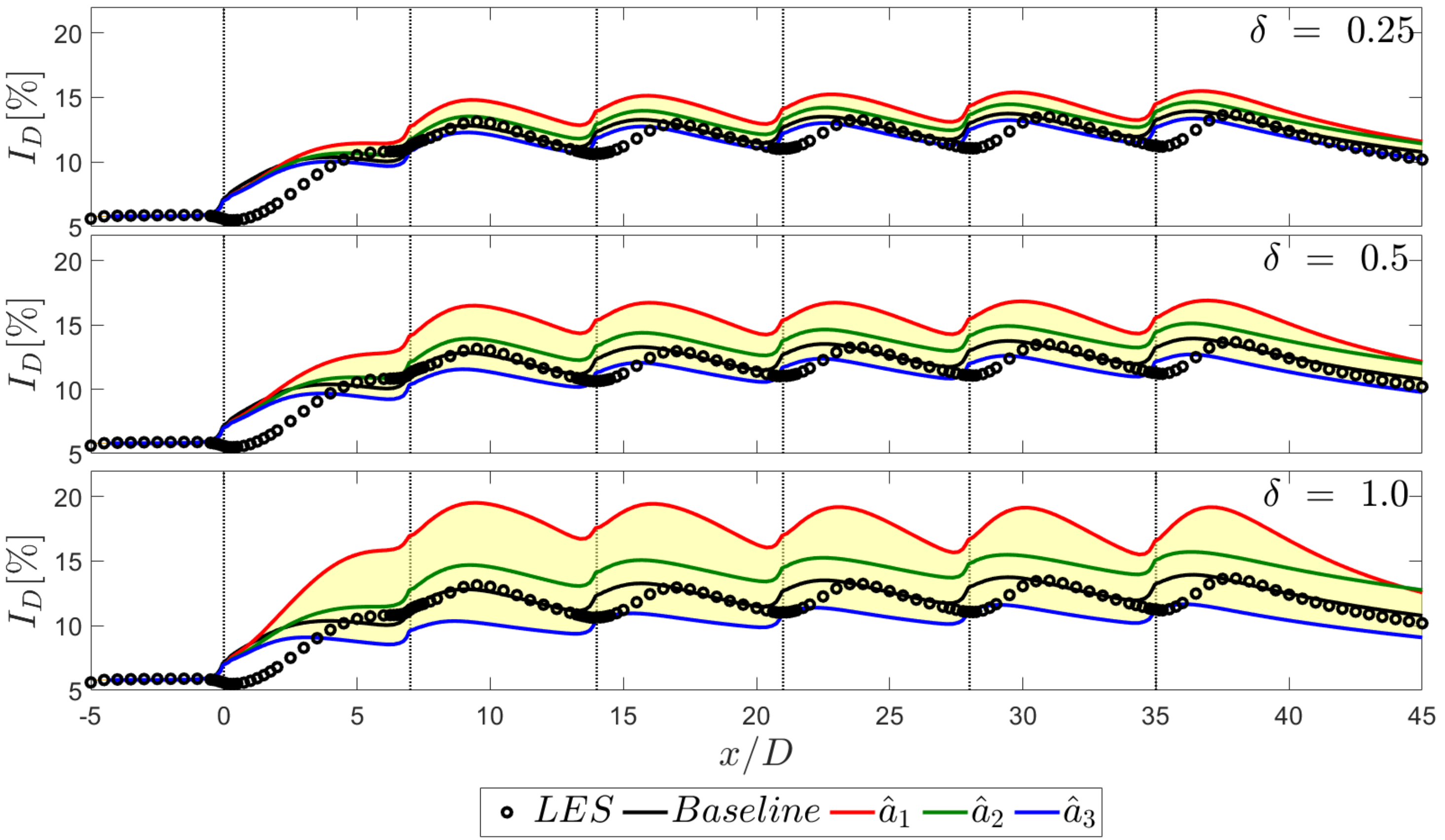}
    \caption{Turbulence intensity, averaged over the rotor area, for LES, Baseline RANS, and perturbed RANS towards one-, two- and three-component turbulence with (from top to the bottom) $\delta = 0.25$, $\delta = 0.5$ and $\delta$ = 1.0.}
    \label{fig:perturbed_average_I}
\end{figure}

\autoref{fig:perturbed_average_I} shows the variation of turbulence intensity, averaged over the rotor, with downstream distance for different values of $\delta$. Consistent with the previous results, perturbation towards one-component and three-component turbulence provides the upper and lower bounds in predicting the turbulence intensity behind the turbines, respectively. Again, it is observed that the higher the amount of perturbation, the wider the bounds. It is also found that all three different perturbation bounds capture the LES data fairly well, except in the vicinity of the turbines, in the near wake region. This might be related to the fact that the baseline model shows poor performance in capturing the magnitude of TKE immediately behind the turbines. 

\begin{figure}
    \centering
    \includegraphics[width=1\linewidth]{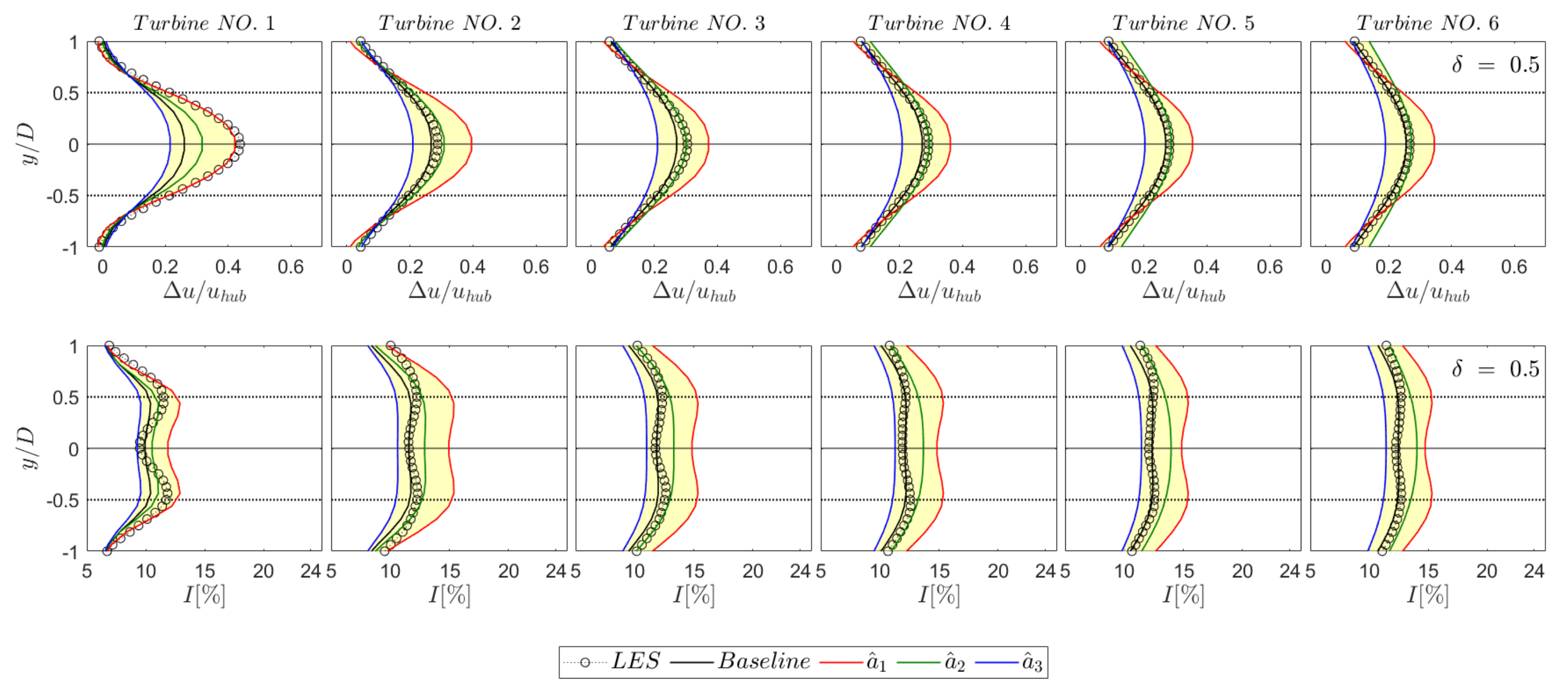}  
    \caption{Lateral profiles of the normalized velocity deficit (top) and turbulence intensity (bottom) at 5D downstream of each turbine at the hub height for LES, Baseline RANS, and perturbed RANS towards one-, two- and three-component turbulence with $\delta = 0.5$.}
    \label{fig:perturbed_NVDI_XD}
\end{figure}

For a more detailed analysis of the effect of perturbation on the QoIs, the lateral profiles of the velocity deficit and turbulence intensity at 5D downwind of each individual turbine are shown in \autoref{fig:perturbed_NVDI_XD} for $\delta=0.5$. As can be seen, the velocity deficit and turbulence intensity obtained from LES are covered relatively well by the bounds obtained from one- and three-component perturbations. 
Consistent with the previous results, perturbation towards one-component and three-component states respectively increases and decreases the velocity deficit in the rotor region. 
Also, as can be seen here, perturbation towards the one-component state reduces the turbulence mixing, yielding stronger the double-peak structure for the turbulence intensity, and a less smeared wake deficit.

\begin{figure}
    \centering
    \includegraphics[width= \linewidth]{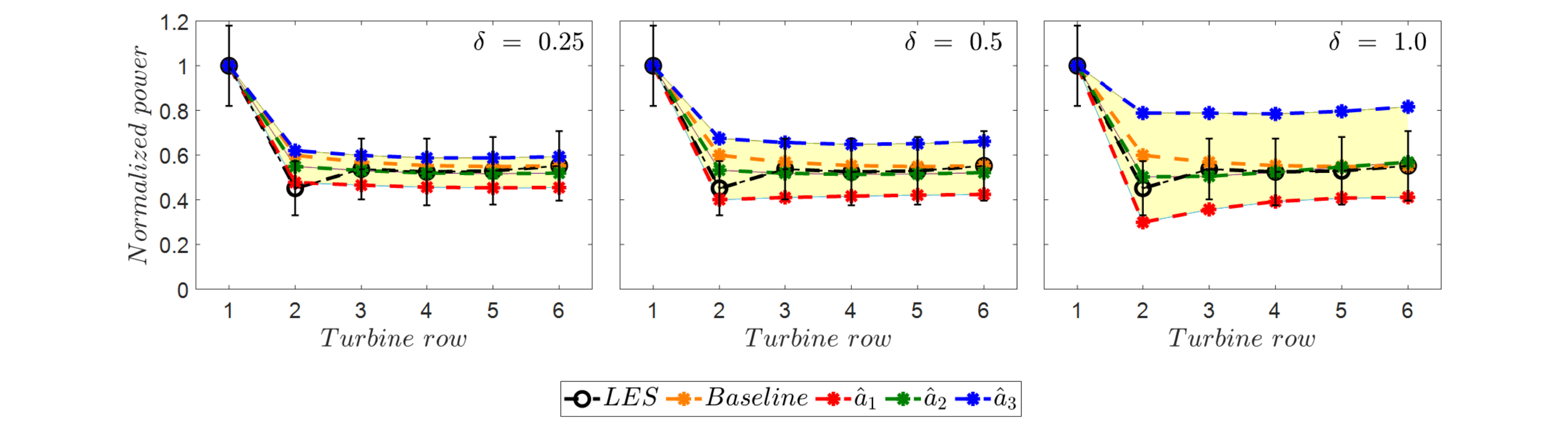}      
    \caption{Normalized power output as a function of turbine row for LES, Baseline RANS, and perturbed RANS towards one-, two- and three-component turbulence with (from left to the right) $\delta=0.25$, $\delta=0.5$ and $\delta$=1.0. The vertical bar shows the standard deviation of the power output obtained from LES. }
    \label{fig:perturbed_power}
\end{figure}

The impact of model-form uncertainty on the predicted power output of the wind farm is shown in \autoref{fig:perturbed_power}. Consistent with the previous results, perturbation towards the one-component turbulence leads to an increase in the velocity deficit and, consequently, a decrease in the power output of the downstream turbines. On the other hand, perturbation toward three-component state enhances the wake recovery and, thus, increases the power output of the waked turbines. It can be seen that when the amount of perturbation value increases, the uncertainty bounds widen. 
It is also found that the model perturbation using the suggested value of $\delta=0.5$ leads to accurate predictions of the power output. 

\begin{figure}
    \centering
    \includegraphics[width= 0.6\linewidth]{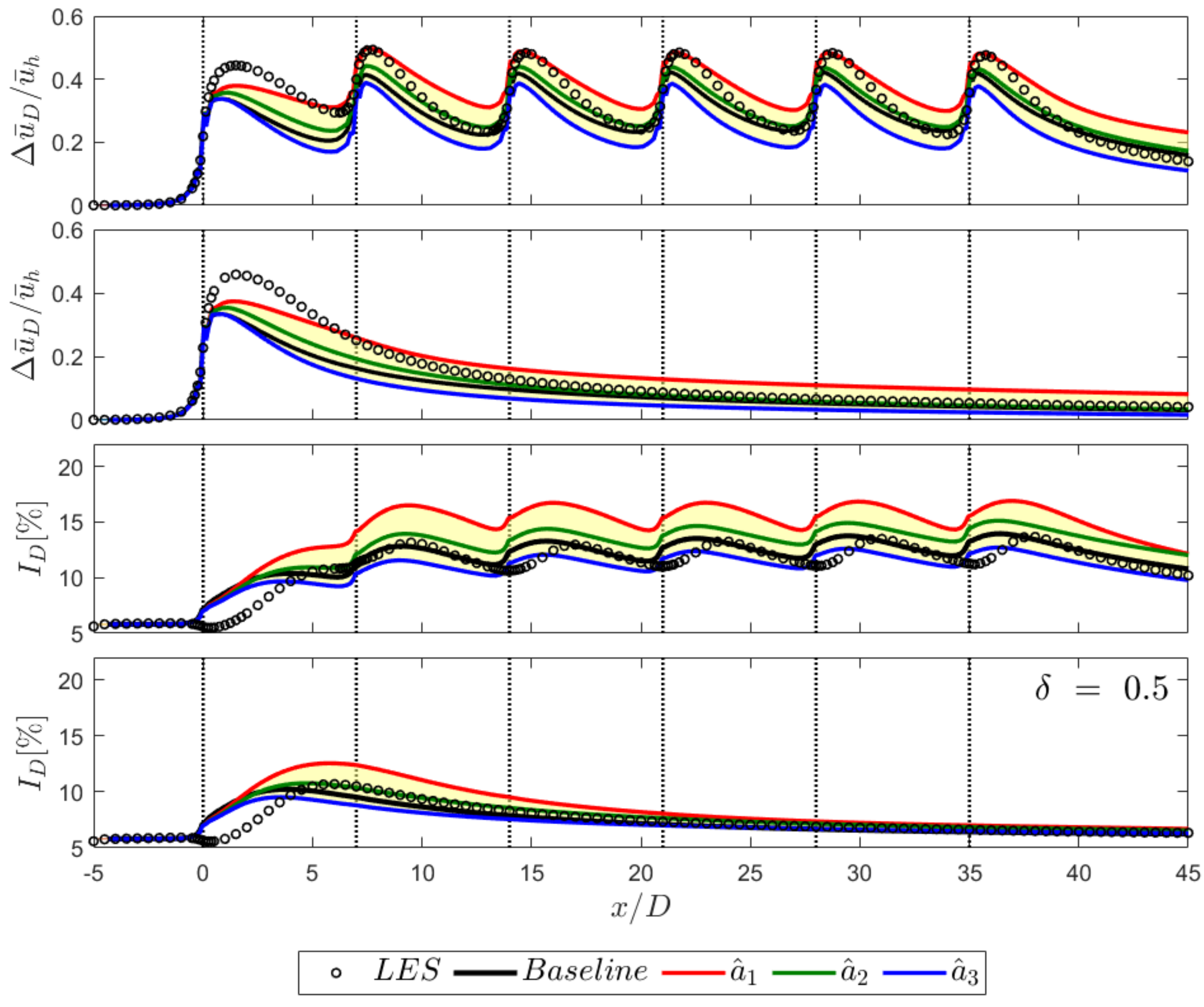}      
    \caption{Normalized velocity deficit and turbulence intensity, averaged over rotor area, for LES, Baseline RANS, and perturbed RANS towards one-, two- and three-component turbulence with $\delta = 0.5$. From top to the bottom: normalized velocity deficit through the wind farm, normalized velocity deficit through the stand-alone turbine, turbulence intensity through the wind farm, and turbulence intensity through the stand-alone turbine.}
    \label{fig:comparison}
\end{figure}

To shed light on the impact of wake interactions on uncertainty bounds, the normalized velocity deficit and turbulence intensity bounds for baseline and single-turbine cases are compared in \autoref{fig:comparison}. As can be seen in this figure, for a single turbine, the uncertainty bound is larger in the region associated with the higher turbulence intensity and decreases with the downstream distance as the wake recovers in the far wake region. However, in the case of the wind farm, the region behind each individual turbine is associated with a high level of shear and turbulence and, consequently, a larger uncertainty bound.

\begin{figure}
    \centering
    \includegraphics[width= \linewidth]{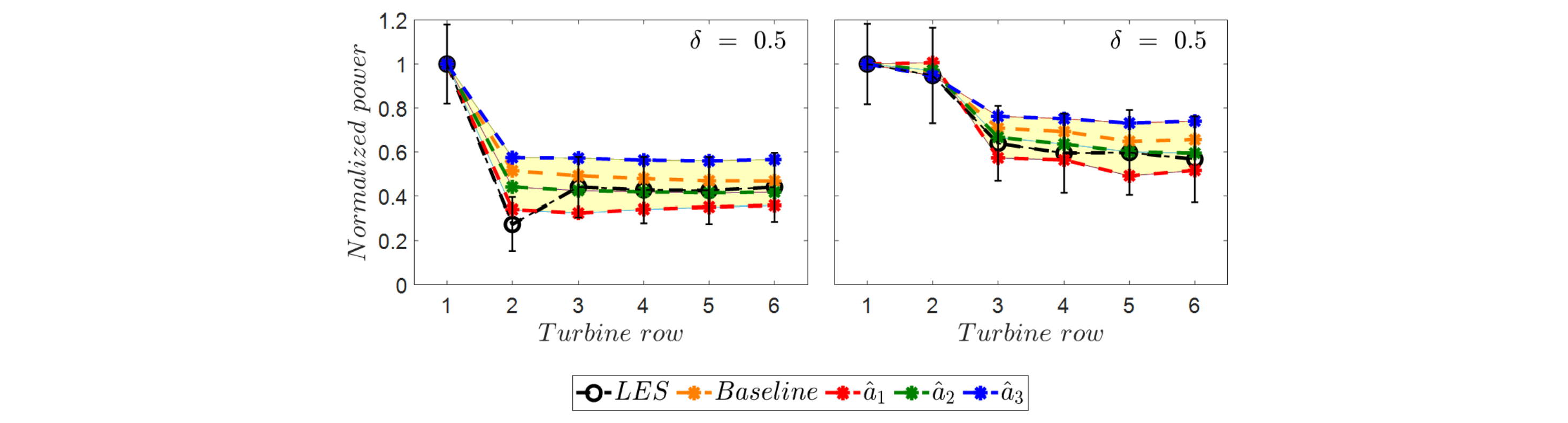}    
    \caption{Normalized power output as a function of turbine row for LES, Baseline RANS, and perturbed RANS towards one-, two- and three-component turbulence with $\delta = 0.5$. Left: aligned configuration with 5D streamwise turbine spacing. Right: staggered configuration with 7D streamwise turbine spacing, while turbine rows 2, 4 and 6 are shifted 1D towards their right side.}
    \label{fig:perturbed_power_partial}
\end{figure}

Finally, in order to further explore the model-form uncertainty on the power losses of wind farms, two additional layouts for the wind farm are considered. In the first configuration, turbines are aligned similar to the baseline case, but the streamwise spacing between them is 5D representing a case with stronger wake interactions. In the second configuration, turbine distances in the streamwise direction are 7D, but the turbine rows 2, 4, and 6 are shifted by 1D to the right side with respect to the most upstream one. The reason for selecting the latter case is to show the performance of the proposed methodology to the case with partial wake interactions. Since the results for velocity deficit and turbulence intensity show a similar trend as the baseline case, and for the sake of brevity, we only show the results related to the power losses. As can be seen in \autoref{fig:perturbed_power_partial}, similar to the baseline configuration, perturbation towards one- and three-component turbulence, with $\delta=0.5$, can bound the LES results for both cases relatively well. These results show the robustness of the method for quantifying the model-form uncertainty in the prediction of power losses of wind farms under both full- and partial-wake conditions.

\section{Summary and concluding remarks} \label{sec:Conclusion}
This study aims to quantify the model-form uncertainty in the RANS simulation of wakes and power losses in wind farms. 
The performance of different RANS models is first tested with the LES data in the prediction of mean velocity deficit, turbulence intensity, and power output from wind turbines. Different RANS models considered in the study are as follow: the standard k-$\varepsilon$ \cite{jones1972prediction, launder1974application}, fp k-$\varepsilon$ \cite{van2018turbulence}, RNG k-$\varepsilon$ \cite{yakhot1992development}, realizable k-$\varepsilon$ \cite{shih1994new}, and SST k-$\omega$ \cite{menter1993zonal}. 
It is found that, the realizable k-$\varepsilon$ model shows a relatively good agreement with the LES data for the above-mentioned quantities, and it is selected as the baseline model for the UQ study. 

In order to quantify the model-form uncertainty in the baseline RANS closure, the state of turbulence is perturbed towards one of the three limiting states; one-component, two-component, and three-component (isotropic) turbulence. In this study, three different values for perturbation amount $\delta$= 0.25 (the least conservative coefficient), 0.5 (obtained from \textit{a priori} studies \cite{hornshoj2021quantifying}), and 1 (extremely conservative approach) are considered.    
Simulation results show that perturbation towards the three-component turbulence leads to a faster wake recovery and reduces the turbulence intensity within the wind farm. Consequently, the power output of the downstream turbines increases when the structure of turbulence is perturbed toward the three-component state. On the other hand, perturbing the turbulence state towards the one-component turbulence leads to an opposite trend: increasing the wake velocity deficit and turbulence intensity and reducing the power output of the waked turbines. Perturbation towards the two-component state does not significantly change the model predictions compared to the baseline model, as it lies within the results corresponding to the one-component and two-component perturbations. It is also found that the value of ${\delta}$ = 0.5 is a suitable choice for creating an acceptable bound covering the LES results.  This is  further confirmed through two additional wind-farm configurations with staggered turbine arrangement and closer turbine positioning. 

In the present study, the focus is placed on perturbation of the shape of the Reynolds stress tensor represented by its eigenvalues.  
Future work can consider the application of Global UQ \cite{huang2021determining} and data-driven machine learning \cite{steiner2020data} to improve the accuracy of the RANS-predicted Reynolds stress tensor in wake-flow simulations.
In addition, the use of non-uniform perturbation magnitude as well as non-uniform regions for perturbations can be further studied (see, e.g., Refs. \cite{heyse2021estimating,steiner2021classifying}).
Future research is also required to address the effect of model-form uncertainties in RANS simulations of wakes and power losses under thermally-stratified conditions and/or over complex terrain.

\section*{Acknowledgements}\label{sec:Acknowledgements}
This study is conducted during the research stay of the first author, as a guest Ph.D. student, at Aarhus University. MA acknowledges the financial support from the Independent Research Fund Denmark (DFF) under the Grant No. 0217-00038B. 
XY acknowledges US Office of Naval Research N000142012315 for financial support.
The authors would like to thank Pourya Forooghi and Frederik W. Johannsen for their valuable support for this work. 

\appendix
\section{Grid convergence study} 
\label{app:appendixA}
\begin{figure}
    \centering
    \includegraphics[width=0.6\linewidth]{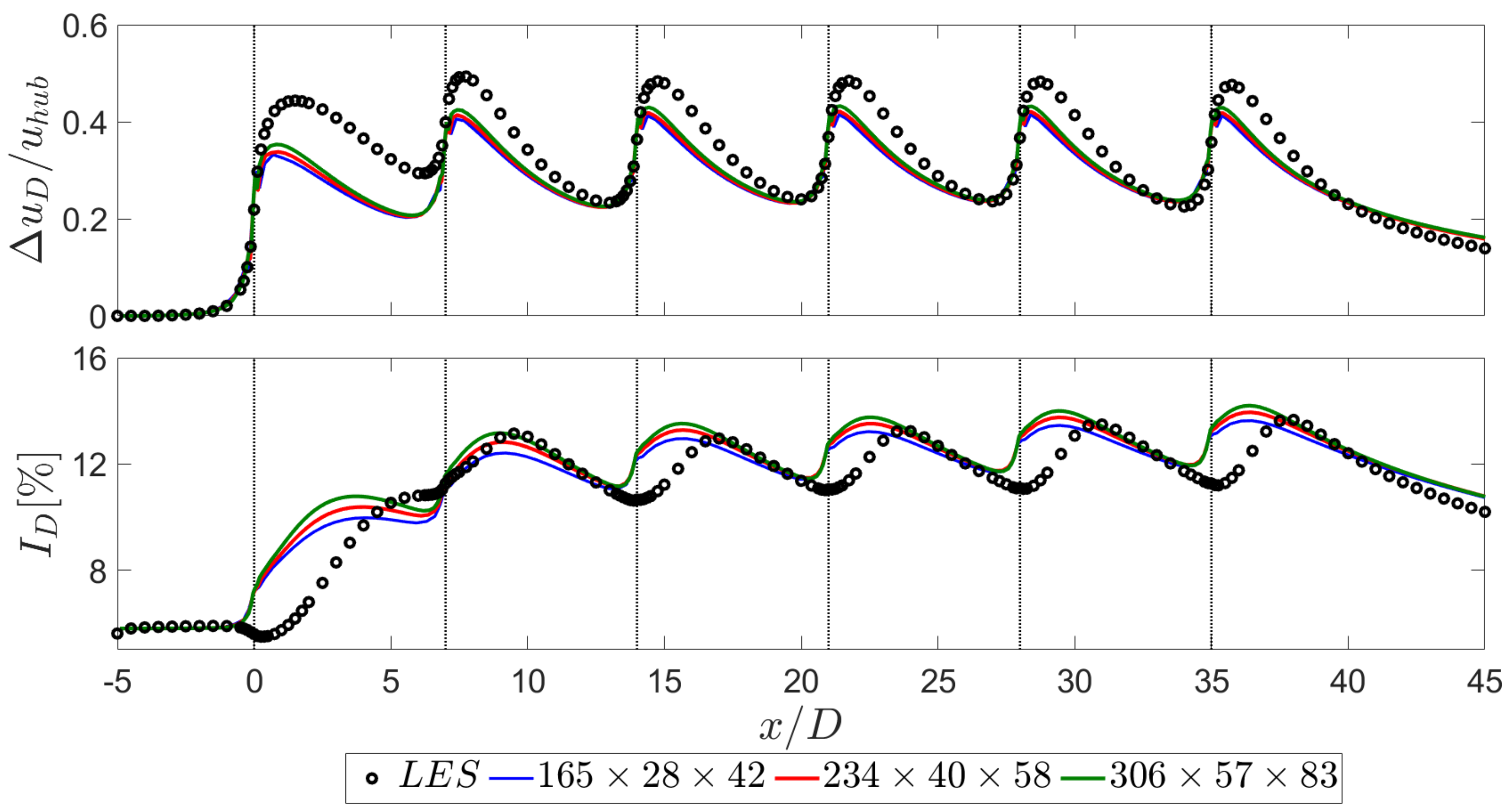}      
    \caption{Normalized velocity deficit (top) and  turbulence intensity (bottom), averaged over rotor area, for different grid resolutions for realizable k-$\varepsilon$ model.}
    \label{fig:grid1}
\end{figure}

\begin{figure}
    \centering
    \includegraphics[width=0.55\linewidth]{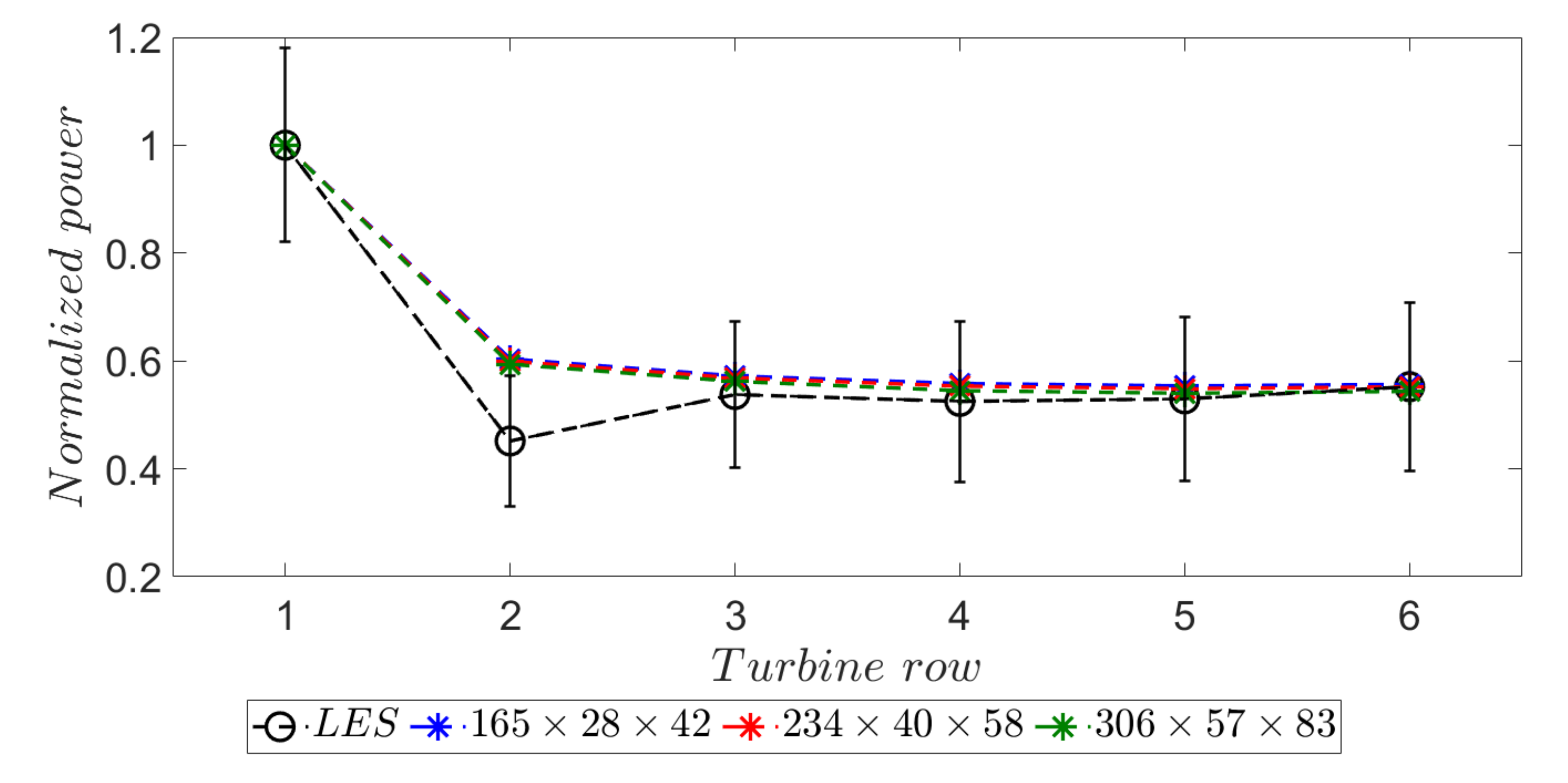}
    \caption{Normalized power output as a function of turbine row for different grid resolutions obtained from the realizable k-$\varepsilon$ model.}
    \label{fig:grid2}
\end{figure}

We assess grid sensitivity by comparing results from three differently-sized grids: $165 \times 28 \times 42$, $234 \times 40 \times 58$ and $306 \times 57 \times 83$. 
\autoref{fig:grid1} shows the variation of the rotor-averaged normalized velocity deficit and turbulence intensity as a function of the downstream distance for different spatial resolutions.
We show results for only the realizable k-$\varepsilon$ model. Results of the other RANS models are similar and are not shown here for brevity.
We see that all the grid resolutions yield fairly similar results for the velocity deficit.
There is a slight difference in the turbulence intensity results behind the first turbine, but the difference becomes small further downstream. 
\autoref{fig:grid2} shows the grid resolution sensitivity result for the normalized power production.
Again, the result does not sensitively depend on the grid resolution. 
Throughout the paper, the results obtained from the medium resolution ($234 \times 40 \times 58$) are presented and discussed. The cell size in the x-direction is uniform away from the turbines (a distance of 1D away). The grid is refined at the turbine location with a grid stretching ratio of 0.5.
A uniform mesh is used in the y-direction.
In the z-direction, we coarsen the grid away from the wall, and the grid size at the top of the domain is three times that at the wall. 

%


\bibliographystyle{elsarticle-num-names}

\end{document}